\documentclass[twoside,fleqn]{article}
\usepackage{espcrc2,latexsym}
\usepackage{epsfig}
\newcommand{\Dsl}{D\!\!\!\!/\,}

\newcommand{\Nf}{N_{\!f}}

\newcommand{\ovr}{\over}

\newcommand{\<}{\langle}
\renewcommand{\>}{\rangle}

\newcommand{\ka}{\kappa}

\newcommand{\rh}{\rho}

\newcommand{\si}{\sigma}

\newcommand{\beq}{\begin{equation}}
\newcommand{\eeq}{\end{equation}}
\newcommand{\bdm}{\begin{displaymath}}
\newcommand{\edm}{\end{displaymath}}
\newcommand{\bea}{\begin{eqnarray}}
\newcommand{\eea}{\end{eqnarray}}
\title{Study of the Leutwyler-Smilga regimes: Lessons for full QCD simulations%
\thanks{Talk given at LAT'00, Bangalore, 17 - 22 Aug 2000.}}

\author{S. D\"urr
\address{Paul Scherrer Institut, Theory Group, 5232 Villigen PSI, Switzerland}%
\thanks{E-mail: {\tt stephan.duerr@psi.ch}}}

\begin{document}

\begin{abstract}
Some key points out of my recent study of the characteristic features of the
small ($x\!\ll\!1$), intermediate ($x\!\simeq\!1$) and large ($x\!\gg\!1$)
Leutwyler-Smilga regimes for QCD-type theories in a finite volume
($x\!=\!V\Sigma m$) are presented, and a few immediate consequences for full
QCD simulations are discussed.
\vspace*{-1mm}
\end{abstract}

\maketitle

\setcounter{footnote}{0}


\section{Motivation}

\begin{figure}
\epsfig{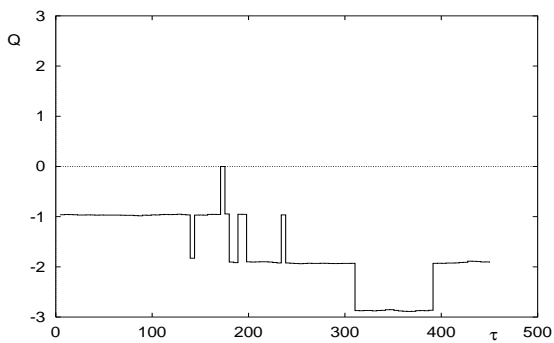}
\vspace*{-13mm}
\caption{Time history of $\nu_{\rm nai}$ in a HMC simu\-lation with a quadruple
of staggered quarks at $\beta\!=\!5.35, m\!=\!0.01$
(figure taken from no.\ 3 in \cite{TopErgodicityHMCstag}).}
\vspace*{-4.6mm}
\end{figure}

Numerical simulations of full QCD
may be hampered by the fact that certain observables show rather long
autocorrelation times.
A well known example is the global topological charge $\nu$ which was
found to develop problems 
for sufficiently small quark masses:
For staggered fermions an insufficient tunneling rate has been observed
both in the symmetric and in the broken phase~\cite{TopErgodicityHMCstag}.
For Wilson type sea-quarks mobility between the topological sectors was found
to be satisfying for $M_\pi/M_\rh\!>\!0.56$, but the same problem emerges if
$\ka$ is tuned sufficiently close to
$\ka_\mathrm{crit}$~\cite{TopErgodicityHMCwils}.
Fig.~1 displays the time-history of a simulation representing the ``worst case
scenario''; the sample is obviously non-ergodic, since the generated
$\nu$-distribution is far from symmetric.

Simply prolonging the simulation time may not be an option; hence one is left
with the question whether the given degree of (non-)ergodicity is tolerable for
practical purposes or whether an ensemble average on the basis of the data
available is afflicted with a sizable systematic error.
It is clear that the answer may depend on the observable one has in mind.
Hence from a physical point of view the question may be rephrased as follows:
Does a given observable on a perfect (full) QCD ensemble --~generated in a box
of given volume $V$ and with $N_f$ sea-quarks of mass $m$~-- depend on the
topological charge $\nu$ of the background ?



\begin{figure*}
\hspace*{1mm}
\epsfig{file=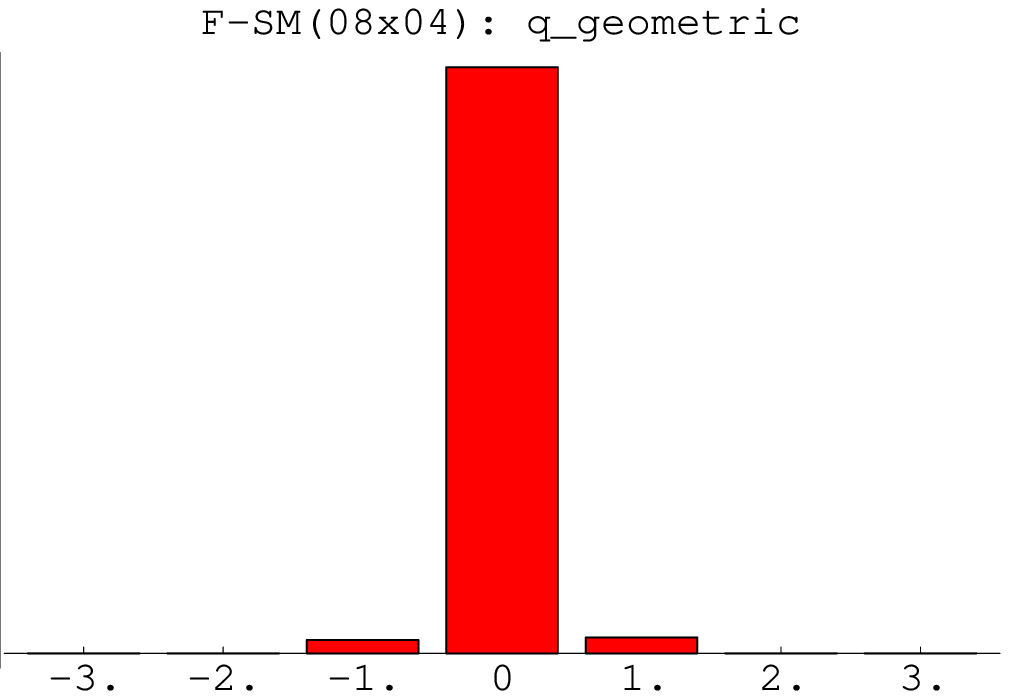,width=5.0cm}
\hspace*{1mm}
\epsfig{file=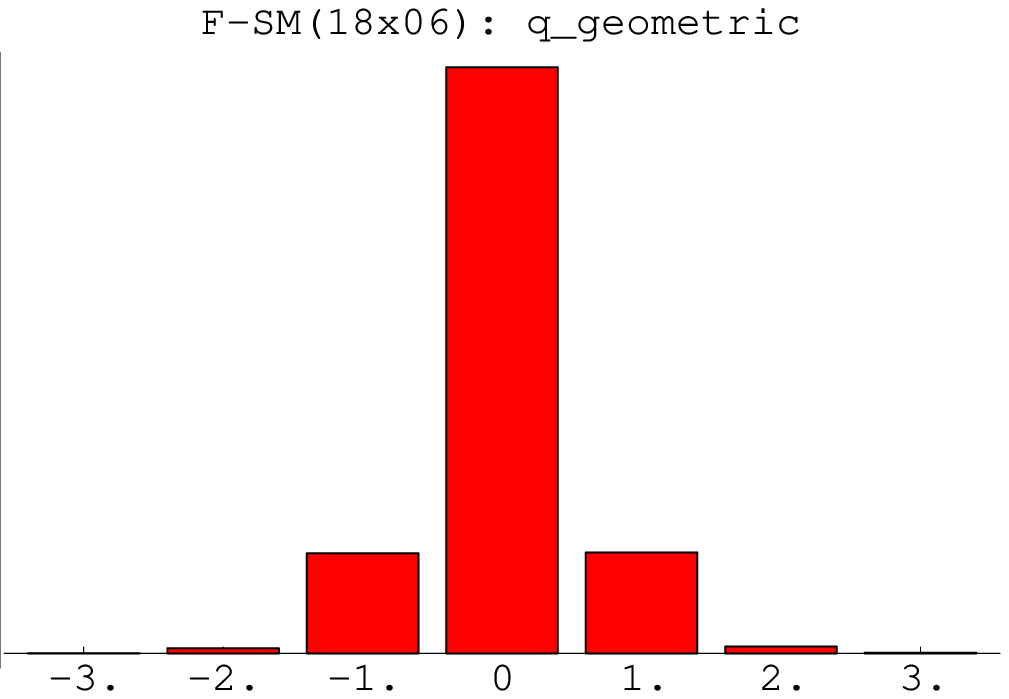,width=5.0cm}
\hspace*{1mm}
\epsfig{file=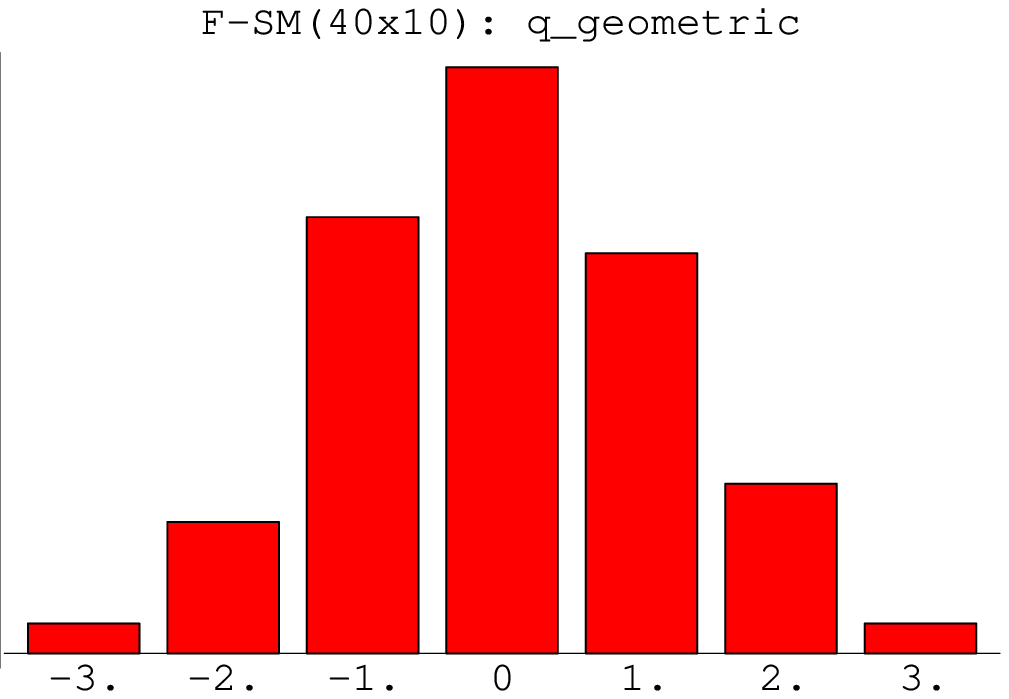,width=5.0cm}
\vspace*{-9mm}
\caption{Distribution of $\nu_{\rm geo}$ in the small (8x4 lattice),
intermediate (18x6) and large (40x10) Leutwyler-Smilga regimes, respectively
[QED(2) data for $\beta\!=\!3.4, m\!=\!0.09, N_{\!f}\!=\!2$].}
\vspace*{-3mm}
\end{figure*}


\section{Leutwyler-Smilga analysis in a nutshell}

For pionic observables, much can be said on purely analytical grounds.
Leutwyler and Smilga (LS) have shown \cite{LeutwylerSmilga} that a clear answer
is possible for two regimes of quark masses and box volumes, both of which
involve the LS-parameter%
\footnote{
$\Sigma\!=\!-\!\lim_{m\to 0} \lim_{V\to\infty} \langle\bar\psi\psi\rangle$
(this order) is the condensate in the chiral limit;
both $\Sigma$ and $m$ are scheme- and scale-dependent, but the combination is
an RG-invariant.}
\beq
x\equiv V\Sigma m
\;,
\label{LSPdef}
\eeq
which indicates whether the box is ``small'' or ``large'' in the sense that it
decides on whether the systems prefers to show symmetry restoration or SSB
phenomena, respectively: For $x\!\ll\!1$ chiral symmetry is effectively
restored and quarks and gluons represent appropriate degrees of freedom.
For $x\!\gg\!1$ the $SU(N_{\!f})_A$ symmetry is quasi-broken, meaning that
--~though the box-volume is formally finite~-- pions represent appropriate
degrees of freedom ($N_{\!f}\!\geq\!2$).
Note that $m$ refers to the mass of the {\em sea\/}-quarks and that $x$ is
(in principle) independent
of the parameter $M_\pi L$.

The net result of the LS-analysis \cite{LeutwylerSmilga} is that for
$x\!\ll\!1$ the partition function is completely dominated by the contribution
from the topologically trivial sector for small enough $m$, since
\beq
Z_\nu \propto m^{N_{\!f}|\nu|}
\;.
\label{LSpeak}
\eeq
In the opposite regime $x\!\gg\!1$ and with the additional condition that the
quark masses are so light that the sea-pion overlaps the box
($M_\pi L\!\ll\!1$), the path-integral in the effective description (requires
$\Lambda_\mathrm{eff}L\!\gg\!1$) is dominated by the constant
mode~\cite{GLthermodynamics}, and Leutwyler and Smilga end up finding that the
partition function is approximately independent of $\nu$ or, more precisely,
that the distribution is~\cite{LeutwylerSmilga}
\beq
Z_\nu \propto e^{-{\nu^2\ovr2\<\nu^2\>}}
\;\;{\rm with}\;\;
\<\nu^2\>\!=\!{V\Sigma m\ovr N_{\!f}}
\;.
\label{LSgaussian}
\eeq



\section{Checking the net charge distribution}

Since the LS-issue is peculiar to the full theory, a preliminary study in a
QCD-type theory seems permissible.
The massive multiflavour Schwinger model (QED(2) with $\Nf\!\geq\!2$) is
supposed to reproduce all qualitative features \cite{Durr}.
I use the Wilson gauge action
$S_{\rm gauge}\!=\!\beta\sum(1\!-\cos\theta_\Box)$ and a pair of
(dynamical) staggered fermions.

I compare the three regimes $x\!\ll\!1$, $x\!\simeq\!1, x\!\gg\!1$ to each
other using three dedicated simulations:
Working with fixed staggered mass $m\!=0.09$ and $\beta\!=\!1/g^2\!=\!3.4$,
the three regimes are represented by the three volumes
$V\!=\!8\!\times\!4, 18\!\times\!6, 40\!\times\!10$.
From this setup the LS-parameter takes the values $x\!\simeq\!0.33,1.12,4.16$,
respectively, and the pion (pseudo-scalar iso-triplet) has a (common) mass
$M_\pi\!=\!0.329$ and therefore a correlation length $\xi_\pi\!=\!3.04$ as to
fit into the box.

A configuration is assigned an index only if the geometric
($\nu_{\rm geo}\!=\!{1\over2\pi}\sum\log U_\Box$) and the
field-theoretic definition ($\nu_{\rm fth}\!=\!\kappa\,\nu_{\rm nai}$,
$\nu_{\rm nai}\!=\!\sum\sin\theta_\Box$,
$\kappa\!\simeq\!1/(1\!-\!\langle S_{\rm gauge}\rangle/\beta V)$, after
rounding to the nearest integer, agree --- for details see \cite{Durr}.

As one can see from Fig.\ 2, the charge distributions found in the three runs
seem to follow a gene\-ral pattern consistent with the LS-prediction:
For $x\!\ll\!1$ the distribution (and therefore the partition function) is
dominated by the contribution from the topologically trivial sector, whereas
for $x\!\gg\!1$ the distribution gets broad and seems compatible with the
gaussian form (\ref{LSgaussian}) with variance $x/N_{\!f}$, i.e.\
$\si^2\!=\!\<\nu^2\>\!=\!x/2$.


\begin{figure*}
\hspace*{1mm}
\epsfig{file=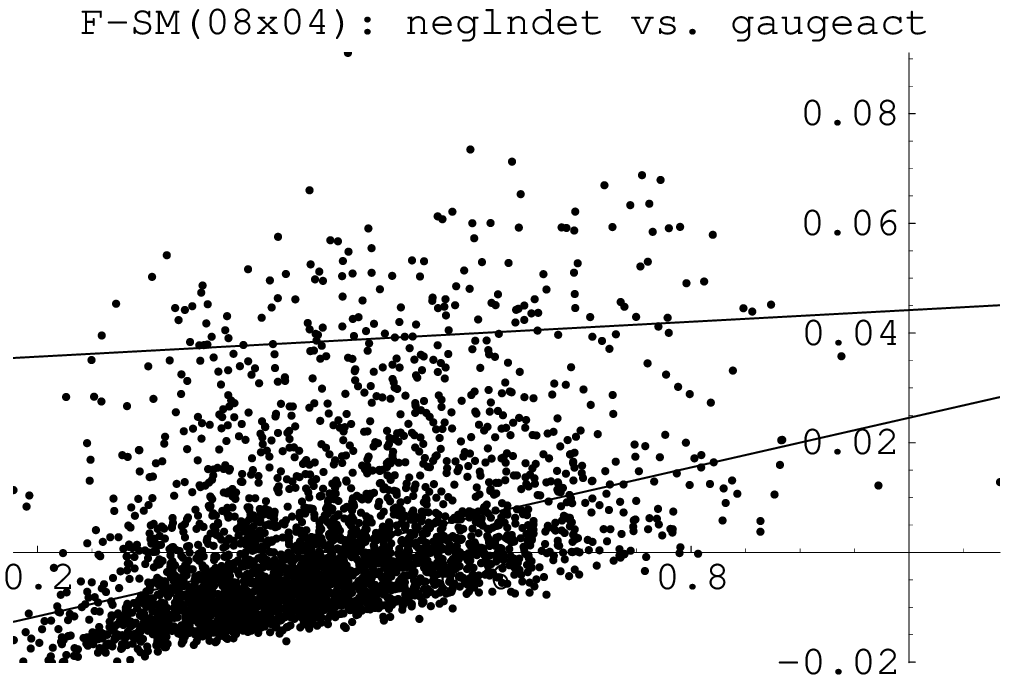,width=4.8cm,height=3.2cm}
\hspace*{1mm}
\epsfig{file=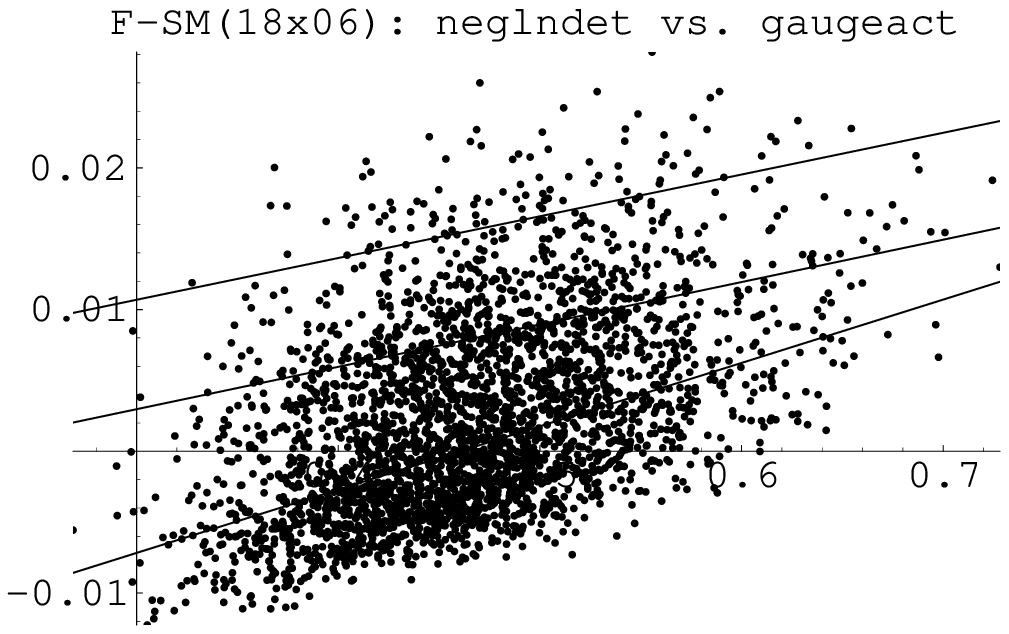,width=5.1cm,height=3.2cm}
\hspace*{1mm}
\epsfig{file=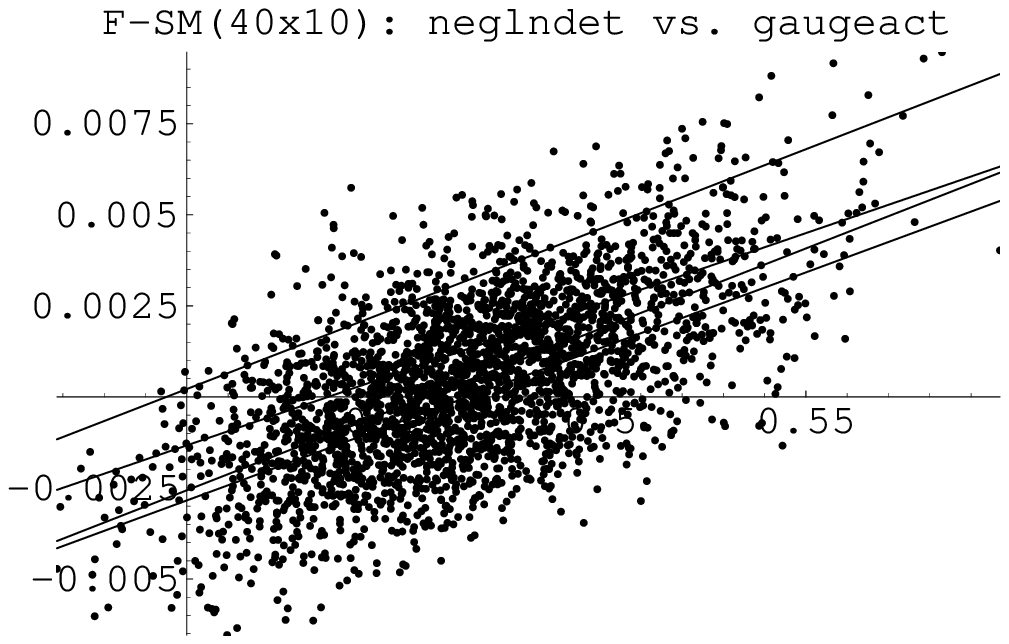,width=5.1cm,height=3.2cm}
\vspace*{-8mm}
\caption{
$S_{\rm fermion}\!=\!-\log(\det(\Dsl\!+\!m))$ (i.e.\ per continuum flavour)
versus $S_{\rm gauge}$ together with the best linear fits for the reasonably
populated topological sectors in the small (8x4 lattice), intermediate (18x6)
and large (40x10) Leutwyler-Smilga regimes, respectively
[QED(2) data for $\beta\!=\!3.4, m\!=\!0.09, N_{\!f}\!=\!2$].\vspace*{-3mm}}
\end{figure*}


\section{Role of the functional determinant}

Since the LS-issue is peculiar to the full (unquenched) theory, an attempt to
understand by which mechanism the three regimes differ from each other may lead
one to investigate how the functional determinant or specifically its
contribution to the total action per continuum-flavour
\beq
S_{\rm fermion}=-\log(\det(D\!\!\!\!/+\!m))+{\rm const}
\label{Sfermion}
\eeq
relates to the contribution from the gauge field and, in addition, how this
might depend on the topological charge of the background.
The idea is thus to study such a relationship sectorally, i.e.\ after the
complete sample has been separated into subsamples with definite absolute
charge $|\nu|$.

From the scatter plots in Fig.\ 3 one notices a (weak) positive correlation
between $S_{\rm fermion}$ and $S_{\rm gauge}$.
This means that the functional determinant acts --~roughly~-- like an effective
renormalization of $\beta$ with a factor bigger than 1.
A key observation is that the correlation improves, if one separates the
sample into subsamples with fixed $|\nu|$ --- for each one of which the best
linear fit for $S_{\rm fermion}$ as a function of $S_{\rm gauge}$ is included.
Quite generally, the linear function for low $|\nu|$ lies below the one(s) for
higher $|\nu|$, and has a larger slope than the latter.
This means that the functional determinant brings --~in general~-- an
{\em overall suppression\/} of higher topological sectors w.r.t.\ lower ones
and a {\em sectorally different renormalization\/} of $\beta$.
There are interesting effects as $x$ varies:
In the small LS-regime both the sectoral dependence of the renormalization
factor and the offset between neighboring sectors are huge, i.e.\ for
$x\!\to\!0$ the functional determinant acts as a constraint to the
topologically trivial sector.
For $x\!\simeq\!1$ both offset and sectoral dependence of the slope are smaller.
In the large LS-regime there is only a minor overall suppression of higher
topological sectors w.r.t\ lower ones, and the effective renormalization of
$\beta$ (and hence, in 4 dimensions, of the physical lattice spacing) seems to
be {\em uniform\/} for all sectors.


\begin{figure*}[t]
\epsfig{file=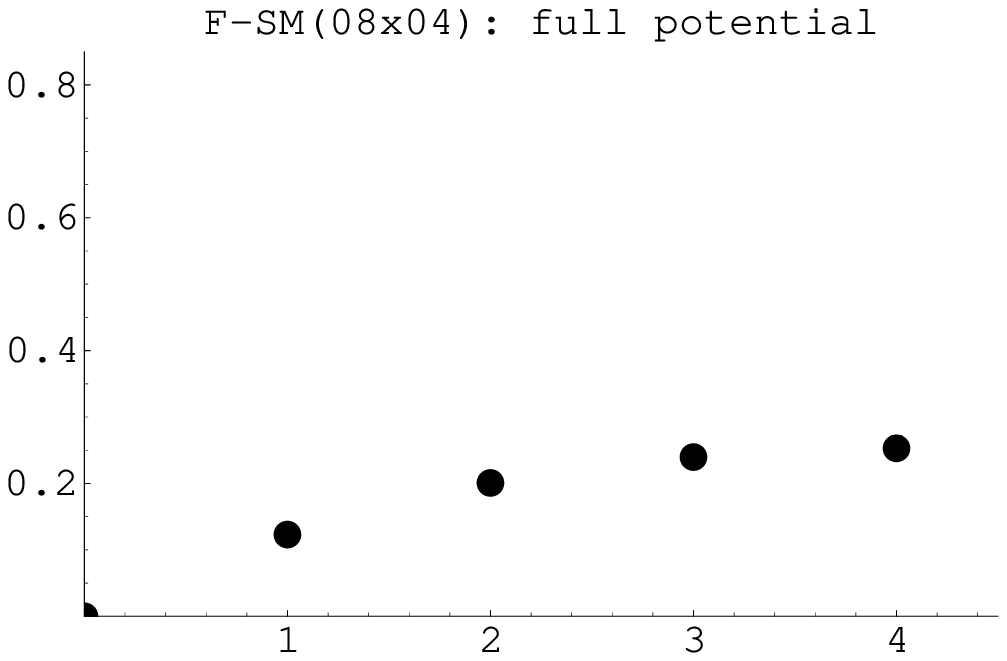,width=5.0cm,height=3.2cm}
\hspace*{1mm}
\epsfig{file=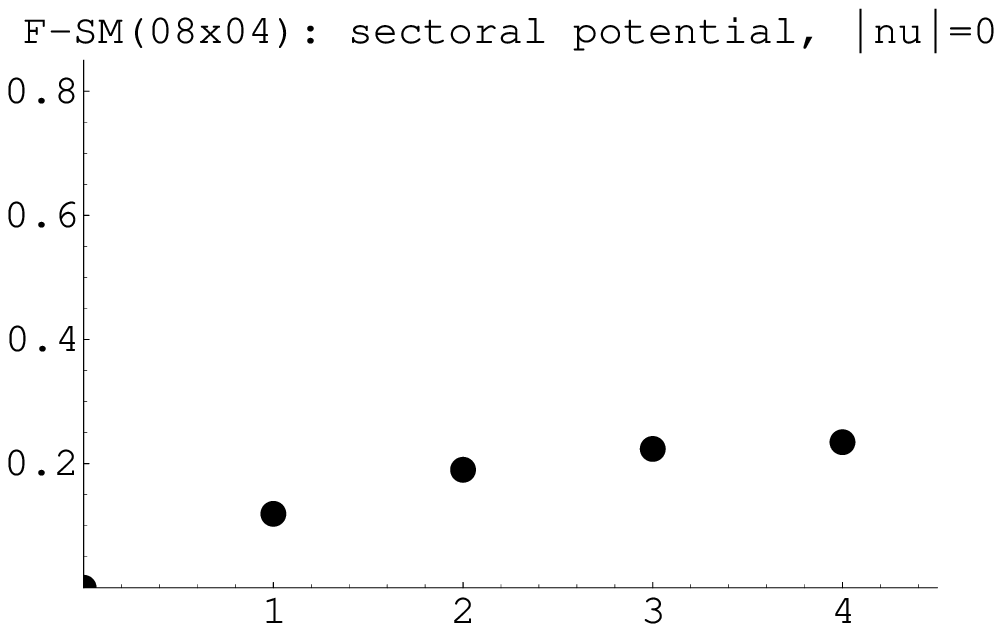,width=5.0cm,height=3.2cm}
\hspace*{1mm}
\epsfig{file=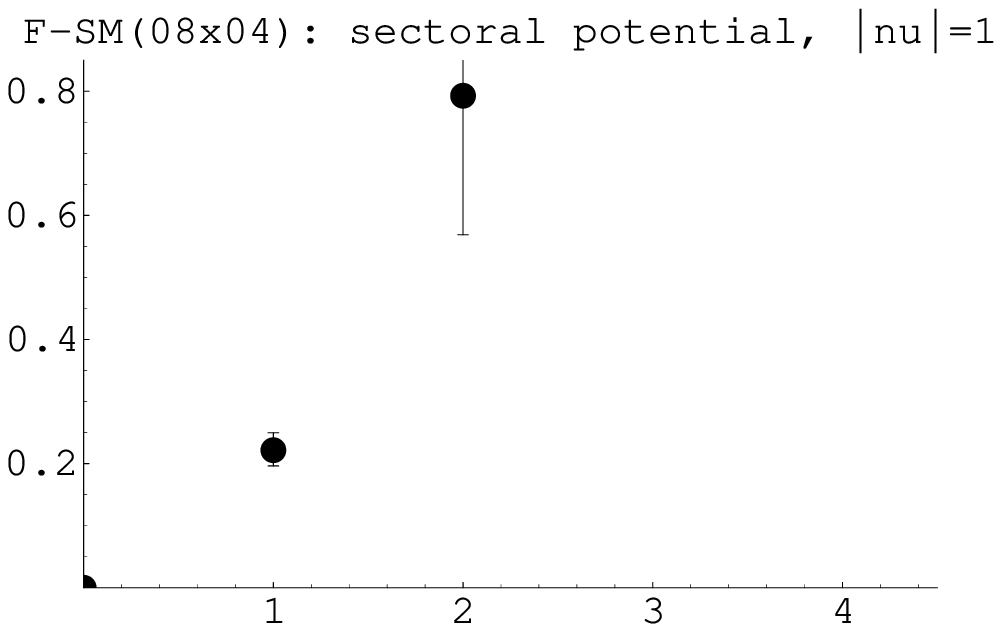,width=5.0cm,height=3.2cm}
\\
\epsfig{file=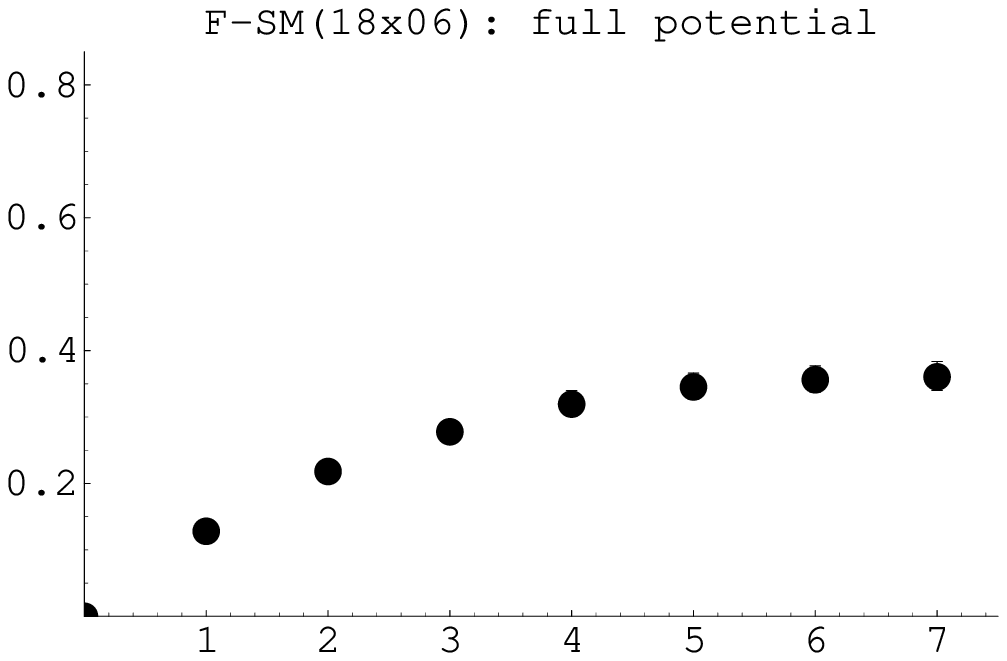,width=5.0cm,height=3.2cm}
\hspace*{1mm}
\epsfig{file=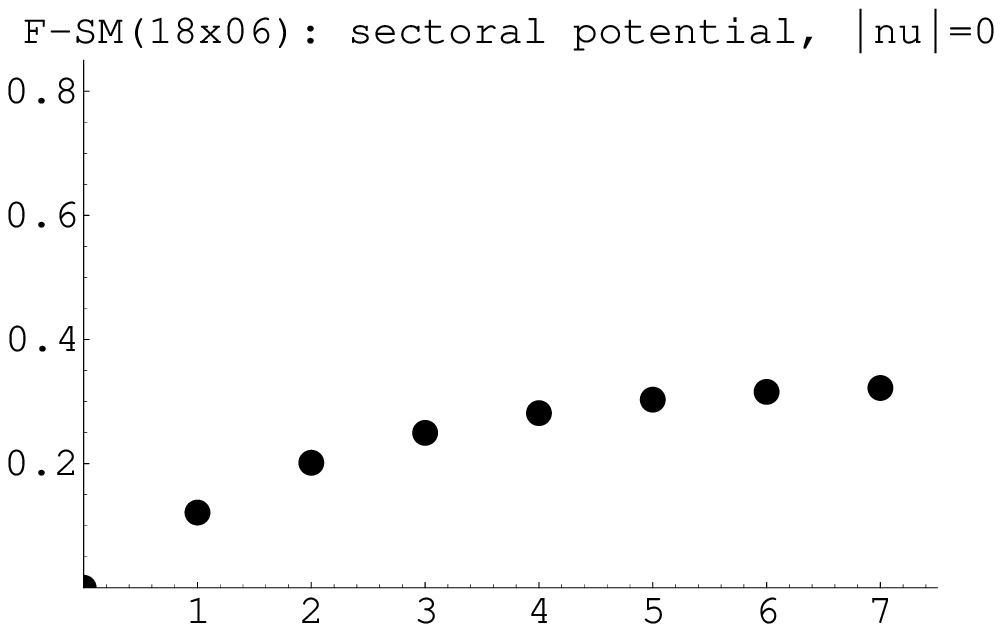,width=5.0cm,height=3.2cm}
\hspace*{1mm}
\epsfig{file=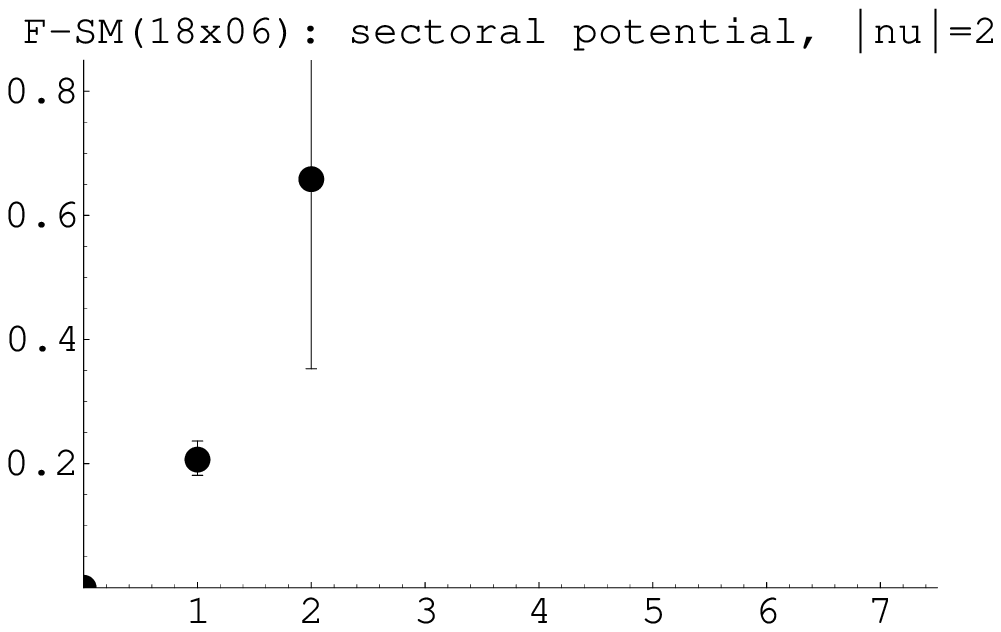,width=5.0cm,height=3.2cm}
\\
\epsfig{file=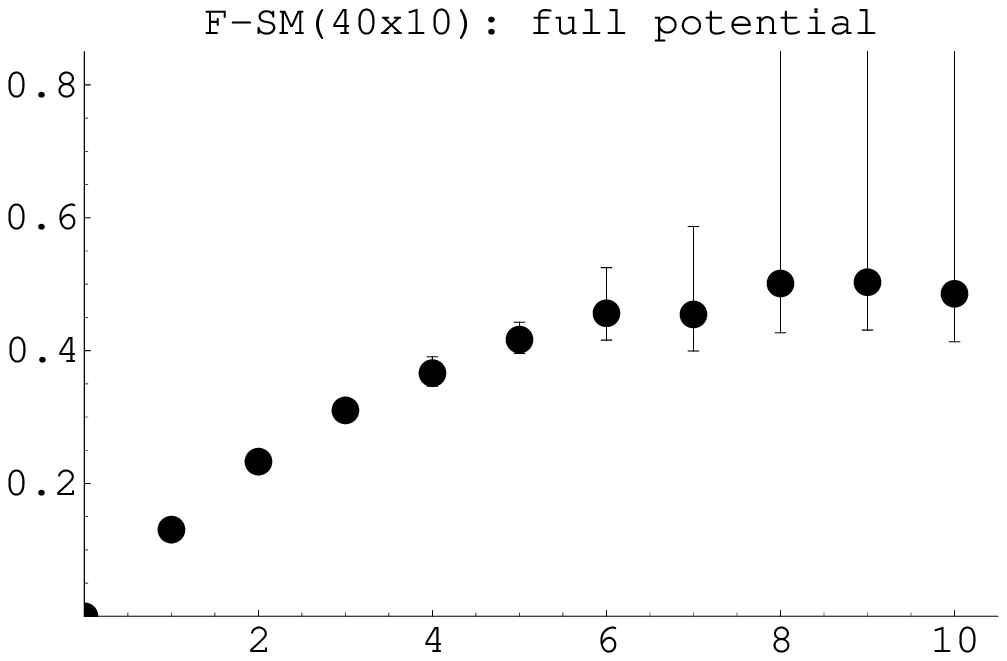,width=5.0cm,height=3.2cm}
\hspace*{1mm}
\epsfig{file=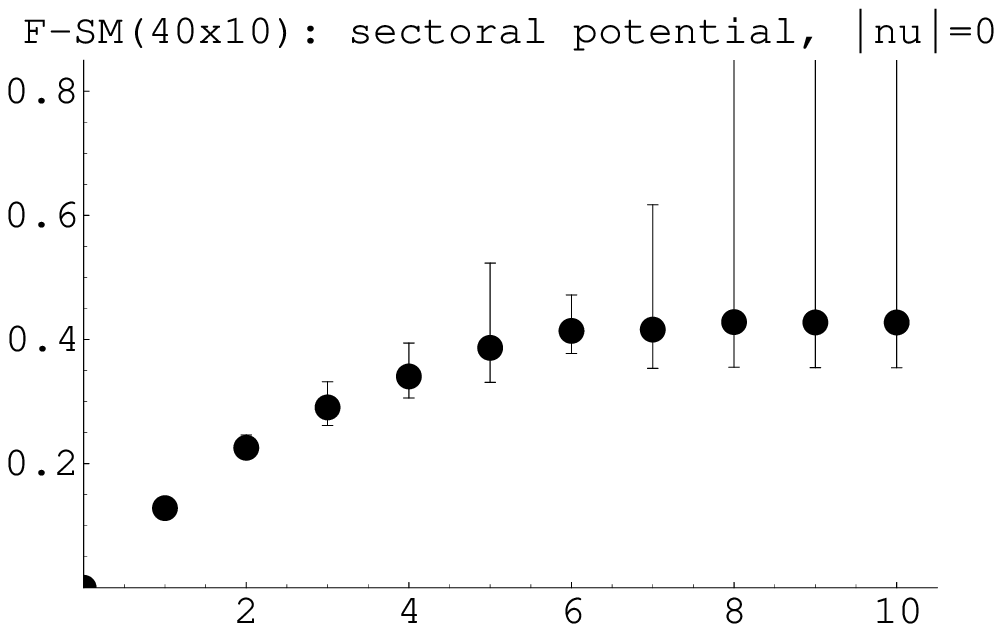,width=5.0cm,height=3.2cm}
\hspace*{1mm}
\epsfig{file=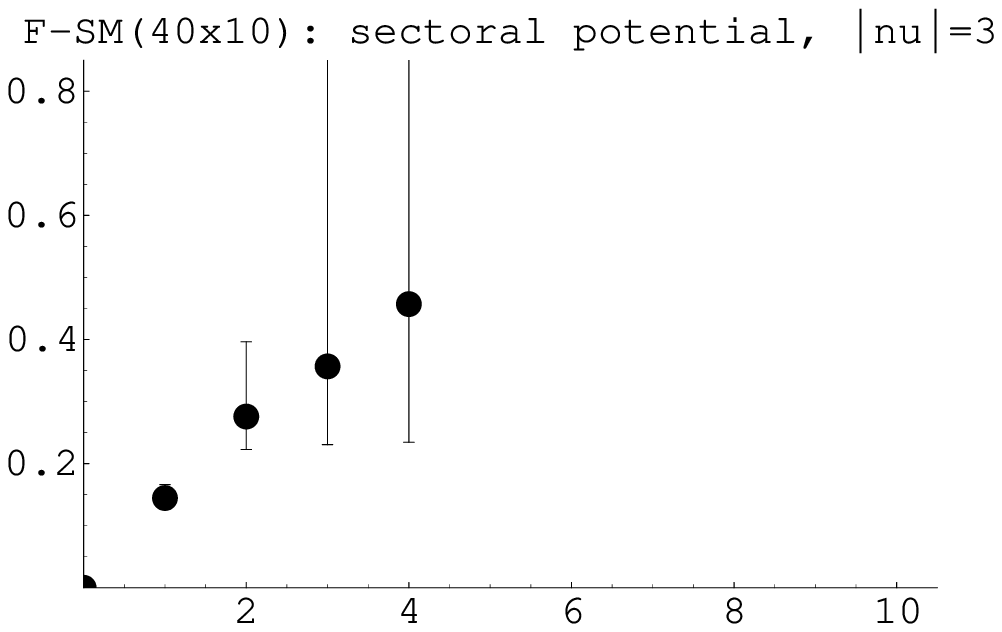,width=5.0cm,height=3.2cm}
\vspace*{-10mm}
\caption{Physical heavy-quark potential (leftmost column) and selected sectoral
versions in the small ($V\!=\!8$x4), intermediate (18x6) and large (40x10)
Leutwyler-Smilga regimes, respectively [QED(2) data].\vspace*{-3mm}}
\end{figure*}


\section{Sectoral heavy-quark potential}

It is instructive to play the same game with a standard observable like the
heavy-quark potential.
As one can see from the leftmost column in Fig.\ 4, the physical potential
(on the unseparated sample) shows, by its curvature, clear evidence of the
dynamical fermions -- the quenched potential would be strictly linear.
The remaining columns show for each LS-regime the sectoral heavy-quark
potentials on the subsets with lowest and highest (reasonably populated)
$|\nu|$.
The sectoral dependence is found to be striking in the regime $x\!\ll\!1$,
where the two subpotentials are clearly inconsistent with each other.
For intermediate $x$, there is still a clear disagreement between the two
extreme-$|\nu|$ subpotentials shown.
In the regime $x\!\gg\!1$, finally, data show good consistency between
neighboring topological sectors (e.g.\ $\nu\!=\!0$ and $\nu\!=\!\pm1$ -- see
\cite{Durr}), but there is no clear evidence of an overall consistency.
In fact, the sectoral potentials for $\nu\!=\!0$ and $\nu\!=\!\pm3$ seem to
indicate that even in the large LS-regime a remnant sectoral dependence might
persists upon generating {\em extreme\/} changes in the absolute value of the
topological charge.
Of course it is possible that in the large-$x$ simulation at hand the value
of $x$ (about 4.16) is not yet large enough for the large-$x$ behaviour to
be fully pronounced, but the rule of thumb seems to be that in the large-$x$
regime standard physical observables may be sensitive to distortions of the
$\nu$-histogram which are large compare to its natural width
$\si\!\simeq\!(x/N_{\!f})^{1/2}$.


\section{Lessons for full QCD simulations}

Each one of the LS-regimes has its characteristic features: For $x\!\ll\!1$
correct sampling w.r.t.\ $\nu$ is easy to achieve; a simple constraint to the
topologically trivial sector proves useful.
For $x\!\simeq\!1$ the situation is much more delicate, since this is the
regime where good sampling w.r.t.\ the topological charge is both crucial and
nontrivial to achieve.
For $x\!\gg\!1$ standard observables (i.e.\ unrelated to the $U(1)_A$-issue)
prove fairly insensitive on the $\nu$-distribution --- unless the latter
suffers from extreme distortions, i.e.\ on a scale which is large compared to
its natural width $\si\!\simeq\!(x/N_{\!f})^{1/2}$.

In spite of the good news that state-of-the-art simulations of full QCD
operate in the large-$x$ regime \cite{DuBu-private},
there is --~in principle~-- a problem with chiral extrapolations.
It derives from the fact that the key role of the LS-parameter is to determine
which one of the limits $m\!\to\!0$ and $V\!\to\!\infty$ is in the ``inner''
position and thus winning:
$x\!\gg\!1$ corresponds to $\lim_{m\to0}\lim_{V\to\infty}$,whereas
$x\!\ll\!1$ implements a situation like $\lim_{V\to\infty}\lim_{m\to0}$.
Therefore one likes to extrapolate from data-points won from a cascade of
simulations for which $x$ {\em increases\/} (or stays at least constant and
$\gg\!1$) as the dynamical quark mass decreases.
Looking at the definition (\ref{LSPdef}), one realizes that $x$ decreases if
a series of simulations on a given grid is used, i.e.\ one is {\em marching in
the wrong direction\/}:
If $\beta$ gets reduced for smaller $m$ so as to keep the lattice spacing
constant (cf.\ Fig.\ 3 and the subsequent discussion), $x$ decreases only in
proportion to $m$; if one works at fixed $\beta$, the situation is of course
worse.
\enlargethispage{1pt}
To correct, one would have to use larger grids for smaller quark masses.
While in current state-of-the-art simulations the box to start
with seems sufficiently large (or the quark mass to end up with heavy enough)
as to prevent this effect from getting numerically important
\cite{DuBu-private}, it is clear that this is a point one may wish to control
as the quest for lighter quark masses goes on.


\end{document}